\documentclass{emulateapj}
\usepackage{amsmath,amssymb,graphicx,natbib}
\newcommand{\Nbody}{\textsl{N}-body\ }
\newcommand{\Mbin}{M_\mathrm{bin}}
\newcommand{\Lbin}{L_\mathrm{bin}}
\newcommand{\Lmin}{L_\mathrm{min}}
\newcommand{\trad}{T_\mathrm{rad}}
\newcommand{\rh}{r_\mathrm{m}}
\begin{document}

\title{The final-parsec problem in nonspherical galaxies revisited}

\author{Eugene Vasiliev\altaffilmark{1,2}} \email{eugvas@lpi.ru}
\author{Fabio Antonini\altaffilmark{3}} \email{antonini@cita.utoronto.ca}
\author{David Merritt\altaffilmark{2}} \email{merritt@astro.rit.edu}
\affil{$^1$Lebedev Physical Institute, Moscow, Russia}
\affil{$^2$School of Physics and Astronomy and Center for Computational Relativity and Gravitation, \protect\\
Rochester Institute of Technology, Rochester, NY 14623, USA}
\affil{$^3$Canadian Institute for Theoretical Astrophysics, University of Toronto, Toronto, Ontario, Canada}
\begin{abstract}
We consider the evolution of supermassive black hole binaries at the center of spherical, 
axisymmetric, and triaxial galaxies, using direct \Nbody integrations as well as 
analytic estimates.
We find that the rates of binary hardening exhibit a significant $N$-dependence
in all the models, at least for $N$ in the investigated range of $10^5 \le N \le 10^6$.
Binary hardening rates are also substantially lower than would be expected 
if the binary ``loss cone'' remained ``full,'' as it would be if the orbits supplying stars
to the binary were being efficiently replenished.
The difference in binary hardening rates between the spherical and nonspherical 
models is less than a factor of two even in the simulations with the largest $N$. 
By studying the orbital populations of our models, we conclude that the rate of 
supply of stars to the binary via draining of centrophilic orbits is indeed expected to
be much lower than the full-loss-cone rate, consistent with our simulations.
We argue that the binary's evolution in the simulations is driven in roughly equal amounts
by collisional and collisionless effects, even at the highest $N$-values
currently accessible.
While binary hardening rates would probably reach a limiting value for large $N$,
our results suggest that we cannot approach that rate with currently available
algorithms and computing hardware.
The extrapolation of results from \Nbody simulations to  real galaxies is therefore
not straightforward, casting doubt on recent claims that triaxiality or axisymmetry 
alone are capable of solving the final-parsec problem in gas-free galaxies.
\end{abstract}

\section{Introduction}
According to large-scale simulations of the clustering of dark matter
in the universe,
the mean time between major mergers of dark halos varies between 
$\sim 0.2$ Gyr at a redshift $z=10$ and $\sim 10^{10}$ yr at $z=1$, 
with a weak dependence on halo mass \citep{Fakhouri2010}.  
Mergers between halo-sized objects are expected to bring the baryonic
components together in a time comparable to the halo coalescence time
\citep{WhiteRees1978,Barnes2001}.
By the same reasoning, 
if each merging galaxy contained a central supermassive black hole (SMBH),
the two SMBHs\ would form a bound system in the merged galaxy
-- a binary SMBH -- shortly after the merger was complete \citep{BegelmanBR1980,Roos1981}.
This idea has received considerable attention because the 
ultimate coalescence of such a binary would
generate an observable outburst of gravitational waves 
\citep[GWs;][]{Thorne1976}.

\citet{BegelmanBR1980} pointed out  a potential
bottleneck in the evolution of binary SMBHs\ toward coalescence.
The binary interacts with nearby stars, ejecting them with velocities comparable
to the binary's orbital velocity. This is the ``gravitational slingshot'' \citep{Saslaw1974}.
But the process is self-limiting, and it is  not a priori clear that the orbits will 
be repopulated in a time shorter than the age of the universe.
This has been called the ``final-parsec problem''  \citep{MilosMerritt2003a}; the name derives
from the fact that the natural separation of a massive binary at the center
of a galaxy is roughly a parsec.

Just as in the case of a {\it single} SMBH at the center of a galaxy,
a binary  SMBH can continue interacting with stars only if the relevant
orbits -- the ``loss-cone'' orbits -- are repopulated.
Repopulation of loss-cone orbits by gravitational encounters -- i.e.,
collisional relaxation -- is well understood in the context of single 
SMBHs in spherical galaxies \citep[e.g.,][]{LightmanShapiro1977,CohnKulsrud1978}.
Collisional effects are less well understood in the 
axisymmetric case \citep{MagorrianTremaine1999,Yu2002,VasilievMerritt2013},
and no such treatment exists yet for triaxial systems.
Additional complications in the case of a binary SMBH include the likely anisotropy
of the initial orbital distribution (just after formation of the hard binary), 
and the fact that the size of the loss region (= binary semimajor axis $a$) changes with time. 

Steady-state loss-cone theory \citep[][chapter 6]{MerrittBook} 
distinguishes between two regimes: 
the empty-loss-cone regime, in which the rate of repopulation of loss-cone orbits is 
slow enough that such orbits are fully depleted; 
and the full-loss-cone regime, which  is reached when 
the encounter rate is so high that it is no longer a limiting factor. 
In the former case, the rate of orbital repopulation is inversely proportional to the 
relaxation time, which itself scales as $\sim N/\ln N$ with the number of stars in the 
galactic nucleus (or particles in the \Nbody simulation), while in the latter case 
it saturates at a value that is essentially $N$-independent. 
The hardening rate of the binary, $s \equiv d(1/a)/dt$, 
is proportional to the rate of repopulation of loss-cone orbits.
Early $N$-body simulations of galaxies containing 
binary SMBHs adopted rather small $N$-values 
\citep{QuinlanHernquist1997,MilosMerritt2001,HemsendorfSS2002} 
and were essentially in the full-loss-cone regime, 
showing little or no dependence of $s$ on $N$.
More recent studies have verified that $s$ drops with $N$ for sufficiently 
high $N$ \citep{MakinoFunato2004,BerczikMS2005,MerrittMS2007}, 
as expected when approaching the empty-loss-cone regime.
Approximate Fokker--Planck models of binary evolution in spherical galaxies 
\citep{MilosMerritt2003b,MerrittMS2007} also predict that the hardening rate of the binary
should scale approximately as $N^{-1}$ for large $N$.

Even in the absence of gravitational encounters, stars can continue to be supplied
to the central binary if their orbital angular momenta are modified by torques
from the nonspherical galaxy.
This ``collisionless'' mode of loss-cone repopulation is essentially independent
of the number of stars (in a galaxy of given size and mass) and can in principle
provide stars to the central binary at high enough rates to ensure coalescence
in a Hubble time \citep{MerrittPoon2004, HolleySigurdsson2006}.
Some recent $N$-body simulations of galaxy mergers suggest in fact that
the rates of binary evolution depend on $N$ weakly, if at all,
a result that the authors have attributed to the nonspherical shapes of the merged galaxies
\citep{KhanJM2011,PretoBBS2011}.
However, other interpretations are possible for these intriguing results, due to
the complex interplay between collisional and collisionless mechanisms.
For instance, the size of the loss region, from which  orbits 
can be driven into the loss cone via {\it collisionless} effects, is much larger in 
nonspherical systems than in spherical ones, hence, it can be more readily repopulated 
by {\it collisional} relaxation. 
Considerations like these suggest that it might be difficult to design an \Nbody 
simulation in which collisional effects would be truly negligible, as they are 
expected to be in many real galaxies.

In this paper we carry out direct \Nbody integrations of binary  evolution 
in spherical, axisymmetric, and triaxial galaxies, constructed initially as 
equilibrium models.
We do not simulate the galaxy merger process, and in this sense, our initial
conditions can be considered less realistic than in some of the studies cited above.
On the other hand, our method lends itself to a more rigorous comparison between
binary evolution rates in galaxies with different geometries. 
We also carry out a much more detailed analysis of the orbital families in our
models, and of the connection between orbital types and the binary evolution
rates seen in the simulations.

\section{Numerical simulations}

\subsection{Model construction}  \label{sec:initial_data}
We considered three series of galaxy models having the same radial density profile but 
different degrees of asymmetry: spherical (S), oblate axisymmetric (A), and triaxial (T). 
The mass distributions were given in each case by a generalization of 
the \cite{Hernquist1990} broken-power-law model: 
\begin{eqnarray}  \label{eq:density_init}
\rho(r) &=& \frac{M_\mathrm{total}}{2\pi abc} \frac{1}{\tilde r\,\left(1+\tilde r\right)^3}  \;, \\
{\tilde r}^2 &\equiv& \left(\frac{x}{a}\right)^2+\left(\frac{y}{b}\right)^2+\left(\frac{z}{c}\right)^2 .
\nonumber
\end{eqnarray}
Axis ratios for the nonspherical models were
 $a:b:c = $1\,:\,1\,:\,0.8 (A) or 1\,:\,0.9\,:\,0.8 (T).
We adopt units such that $abc=M_\mathrm{total}=1$.
The potential of a central point mass, $-GM_\mathrm{bin}/r$, 
representing the massive binary was added to the self-consistent potential of the stars;
we consider a single value for $\Mbin=10^{-2}$, which is typical of previous studies. 
This is somewhat larger than the typical ratio $\sim 10^{-3}$ of SMBH mass to 
galaxy mass \citep[e.g.][]{MarconiHunt2003}; 
however, since most stars that interact with the binary come from radii smaller 
than the scale radius of the model, restricting our consideration to the central
parts of a galaxy is not likely to strongly affect the results.
The distribution function for the spherical models was created 
using Eddington's inversion formula \cite[][Equation~(3.47)]{MerrittBook},
which gives the unique, isotropic $f(E)$ corresponding to a specified
potential and density.
The nonspherical \{A, T\} models were constructed by \citet{Schwarzschild1979} orbit 
superposition method,
using the publically available \textsl{SMILE} software \citep{Vasiliev2013} and $10^5$ orbits.
The nonspherical models are intrinsically anisotropic in velocity space, 
but we imposed ``maximal isotropy'' by requiring  the velocity dispersion 
in the radial direction to equal one-half the sum of the velocity dispersions in the
two transverse directions (no additional constraint was placed on the latter).

Monte Carlo realizations of each model were constructed for values of $N$ in the range
$8\times 10^3\le N \le 10^6$.
We created several, independent realizations for each $N$ 
(four for $N\le 125$K, two for $N=250$K and 500K, one for $N=1000$K) 
because simulations with small $N$ exhibit considerable scatter in the rate of evolution
\cite[e.g.,][]{MerrittMS2007}.
Unless otherwise specified, data plotted in the figures consist of values averaged 
over the multiple realizations. 

We then replaced the single massive particle by two SMBH particles each of mass 
$\Mbin/2=5\times 10^{-3}$ located symmetrically about the origin at 
$x=\pm 0.1$ \citep[as in][]{MerrittMS2007}. 
The initial separation was slightly larger than the radius of influence $\rh$, defined as
the radius containing a stellar mass equal to $2\Mbin$; initially $\rh\approx 0.15$ 
and $\rh$ increases to about 0.2 as the central density drops.
The initial velocities of the SMBH particles were set to $0.31$ in model units, 
corresponding to a circular orbit in the $x-y$ plane. 
In the course of the evolution, the eccentricity of the binary was found to
remain low for $N>10^5$, although it became somewhat larger 
for smaller $N$, with a large scatter between realizations; 
the average eccentricity was $\approx 0.2(N/10^5)^{-1/2}$. 

We might also have placed just one SMBH particle at the origin and allowed the 
other to spiral in, or placed the two SMBHs symmetrically into an equilibrium model 
created without a central mass.
We experimented with these and other configurations but found that the choice of 
initial placement affected only the initial stage of evolution, 
not the behavior at the hard binary stage.

We verified that the models so constructed were in equilibrium by following their evolution 
for 20 time units (with a single central point mass) using the $N$-body code
described below.
No discernible evolution of the density profile or the model shapes was observed.

\subsection{Parameters of the \Nbody integrations}  \label{sec:simulation_params}
\begin{figure*}
$$\includegraphics{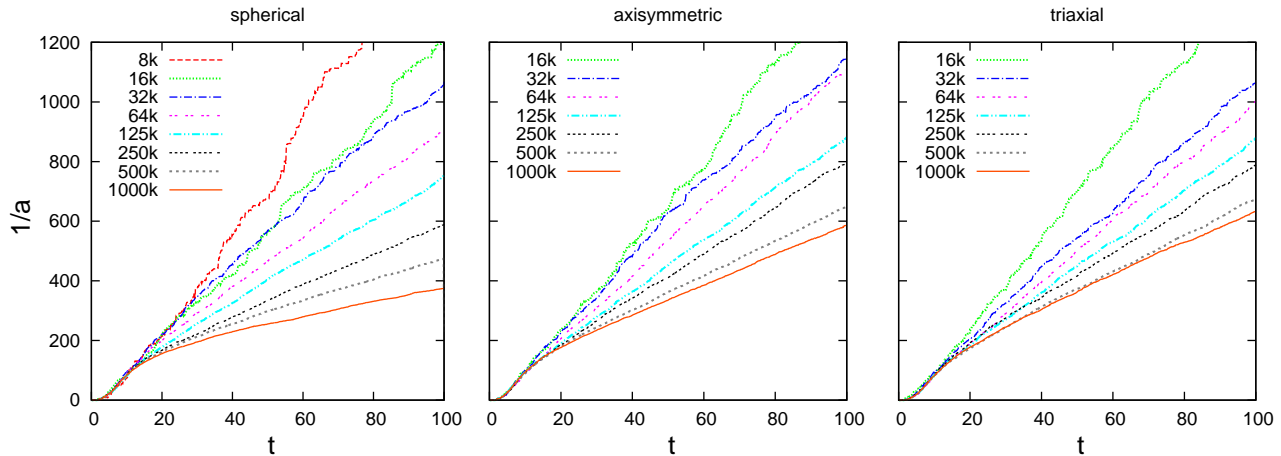}$$
\caption{
Evolution of binary hardness, $1/a$, as a function of time, for three series of models: 
spherical (left), axisymmetric (middle), and triaxial (right). Different curves are 
for models with $N$ varying from $8\times10^3$ to $10^6$ (from top to bottom),
averaged over several realizations with the same $N$, as discussed in the text.
} \label{fig:hardness}
\end{figure*}

We used the direct \Nbody integrator $\phi$GRAPEch \citep{HarfstGMM2008}
to follow the evolution of the massive binary.
This code combines hardware-accelerated
computation of pairwise interparticle forces (using the \textsl{Sapporo} library 
\citep{GaburovHP2009}, which emulates the GRAPE interface utilizing GPU boards)
with a high-accuracy chain regularization algorithm to follow the dynamical interactions of 
field stars with the two SMBH particles. 
The chain radius was set to $4\times 10^{-3}$ length units. 
In the present implementation, there could be only one chain 
which necessarily includes the first SMBH particle. Hence, in the early stages of evolution 
close approaches of field stars with the second SMBH particle were not regularized, which 
in principle might have led to the accumulation of errors; nevertheless, the relative error 
in total energy was typically $\sim 10^{-4}$ for the accuracy parameter $\eta=0.01$ and even 
much smaller in the later stages of evolution ($t\ge 30-40$ time units), when both 
SMBHs were  included in the chain.

We used zero softening for interactions in the chain and set a very small softening length 
$\epsilon=10^{-6}$ outside the chain to prevent energy errors at the early integration stages. 
This is a much smaller softening length than 
the values typically used in other studies, and also much smaller than the distance of strong 
deflection for encounters between field stars; hence, we are guaranteed not to change the
effective value of the Coulomb logarithm or the rate of relaxation. 
We checked this by repeating some simulations with 
$\epsilon=0$ at the late stages and verifying that there 
was no substantial difference in the binary hardening rate.
Our experiments indicated that a larger value of softening length ($\epsilon\ge 10^{-5}$) 
decreases the hardening rate and reduces the difference between simulations with different 
$N$, as well as between S, A, and T models.

\subsection{Results}  \label{sec:simulation_results}
\begin{figure}
$$\includegraphics{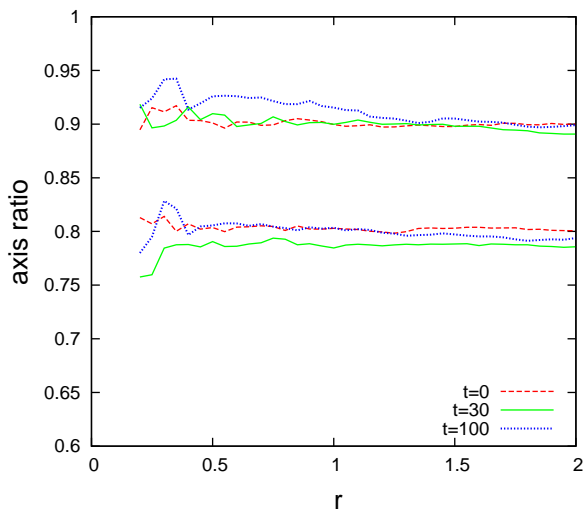}$$
\caption{
Axis ratios of the triaxial model with $N=10^6$ as a function of radius: 
initially (red), after formation of a hard binary ($t=30$, green), and still further into 
the evolution ($t=100$, blue). The model remains triaxial and close to its original shape. 
} \label{fig:axis_ratio}
\end{figure}

\begin{figure}
$$\includegraphics{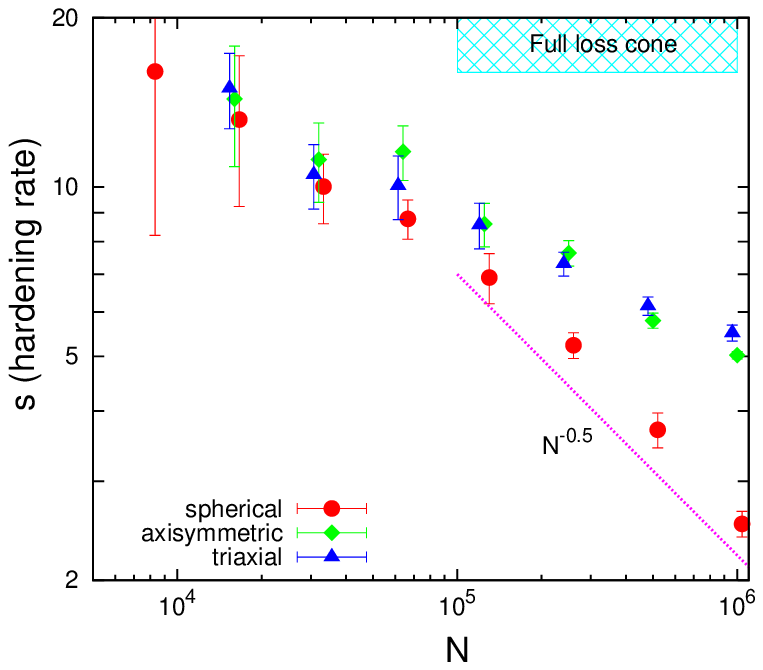}$$
$$\includegraphics{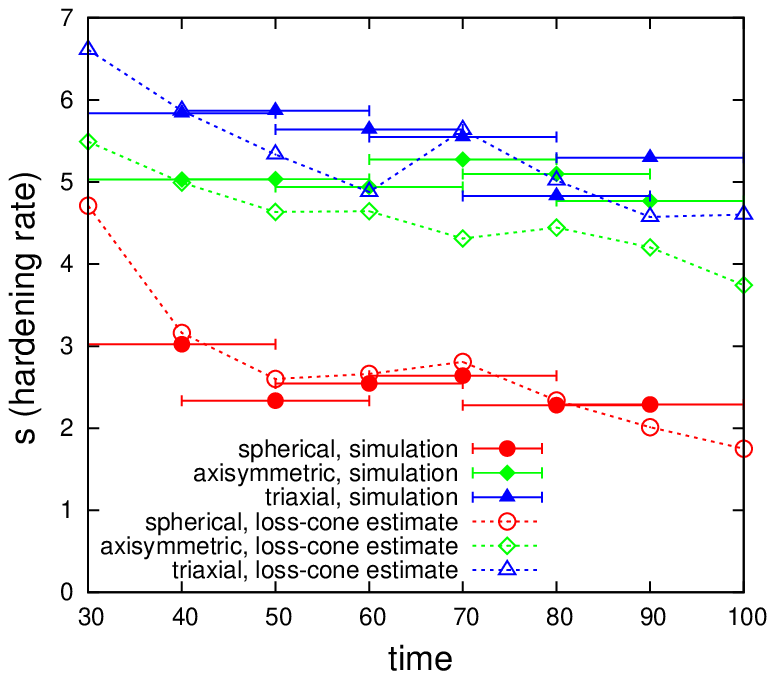}$$
\caption{
\textit{Top panel:}
hardening rates $s\equiv (d/dt)\left(1/a\right)$ as a function of $N$ for three series 
of models (red, spherical; green, axisymmetric; blue, triaxial), 
measured on the interval $30\le t \le 100$ (at the hard binary stage). 
Error bars reflect the variance seen in different realizations of the same model. 
The hardening rate for a full loss cone of the binary $s=18\pm 2$ is marked by the 
hatched region on the plot. \protect\\
\textit{Bottom panel:}
hardening rates as a function of time for $N=10^6$ models (colors are the same as 
in the top panel). 
Filled symbols with horizontal error bars show the estimate of the slope of $1/a(t)$ 
on the corresponding interval of time; dashed lines with open symbols show the 
estimates from the loss-cone population (Equation~\ref{eq:hardening_rate_nbody}) 
at corresponding moments of time. The two estimates agree quite well and have 
a rather moderate variation over time, with a weak tendency to decline.
} \label{fig:hardening_rate}
\end{figure}

The models were evolved for 100 time units, 
with the final value of the binary semimajor axis, $a$, reaching 
$a\lesssim (1-2.5)\times 10^{-3}$ length units depending on $N$ and on model type 
(Figure~\ref{fig:hardness}). 
The elapsed time until formation of a ``hard binary,'' $a\lesssim a_\mathrm{hard}$,
was roughly $t=20$; we define $a_\mathrm{hard}$ in the standard way as
\begin{equation}
a_\mathrm{hard} \equiv \frac{\mu}{\Mbin}\frac{\rh}{4} \;, \quad 
\mu \equiv \frac{M_1 M_2}{M_1+M_2} 
\end{equation}
\cite[][Equation~(8.71)]{MerrittBook},
with $\mu$ the reduced mass of the binary.
For our models, $a_\mathrm{hard}= \rh/16 \approx 10^{-2}$.

Formation of the hard binary was accompanied by a substantial change in the 
density profile and in the distribution of particle energies in the models. 
Roughly speaking, the original $\rho\sim r^{-1}$ cusp was replaced by a shallower, 
$\rho\sim r^{-1/2}$ density profile inside $\rh$,
corresponding to a mass deficit \citep{MilosMRB2002} of order $\Mbin$. 
The change in the energy distribution was more dramatic; 
almost no particles remained for energies $|E|\gtrsim |\Phi_0|\approx 0.8$, 
where $\Phi_0$ is the depth of the potential well due to the stars alone.
In other words, there are  almost no particles left that would be bound to the massive binary. 
In the course of subsequent evolution, the density profile and distribution function changed 
only modestly, the changes being spread over a much wider range of radii. 
Figure~\ref{fig:axis_ratio} shows that the axis ratios of the models at the hard binary stage, 
determined iteratively from the tensor of inertia of \Nbody snapshots \citep{Katz1991}, 
were quite close to their values at $t=0$. In other words, the binary has not 
destroyed the large-scale flattening or triaxiality, despite introducing some changes within 
the influence radius.

For $t\gtrsim30$, the binary hardening rate $s\equiv d(1/a)/dt$ was observed to be
almost unchanging in each simulation. 
The hardening rate was computed as the slope of the inverse semimajor axis plotted 
as a function of time, calculated for several overlapping intervals of 20 time units 
on the interval $30\le t \le 100$ and averaged to get an estimate of scatter.
We also averaged the results over the different realizations with the same $N$, 
adding the scatter to the error bars.
Figure~\ref{fig:hardening_rate}, top panel, shows that for low $N$ ($\lesssim 3\times10^4$) 
there is little difference between the three models and a weak dependence on $N$, 
consistent with earlier studies that also used small particle numbers 
\citep{QuinlanHernquist1997,MilosMerritt2001,ChatterjeeHL2003}. 
For $N\gtrsim10^5$ the spherical model demonstrates a clear dependence of 
hardening rate on $N$, approximately as $s \propto N^{-0.5}$, 
again in agreement with earlier studies
\citep{MakinoFunato2004,BerczikMS2005},
and consistent with theoretical models of collisional loss-cone refilling
\citep{MerrittMS2007}. 

Hardening rates in the nonspherical models A and T are somewhat larger than in the
spherical model, and quite close to each other, the difference appearing only for the $N=10^6$ 
model. But -- contrary to our initial expectations, based on the galaxy merger studies
cited in Section 1 -- the hardening rates still exhibit a 
clear $N$-dependence at large $N$, suggesting a continued role for collisional
orbital repopulation in these models.

\section{Analysis of hardening rates}

\subsection{Predictions from scattering experiments}  \label{sec:scattering_estimates}
To understand the binary evolution found in the simulations, we begin by considering
the hardening rate due to the interaction between the binary and incoming stars 
in the \Nbody models. 
Following \cite{Hills1983}, we define the dimensionless coefficient $C$ describing 
the energy exchange in one interaction between a ``field'' star of mass $m_\star$
and a massive binary:
\begin{equation}  \label{eq:coefC}
C \equiv \frac{\Mbin}{2m_\star}\frac{\Delta E_\mathrm{bin}}{E_\mathrm{bin}} \;, 
\quad E_\mathrm{bin} \equiv -\frac{G\Mbin}{2a} .
\end{equation}

The strength of an interaction can also be expressed in terms of the velocity 
at infinity, $v$, and impact parameter, $b$, of the incoming star. 
In the hard-binary limit, $a\ll a_\mathrm{hard}$, the scattering outcome 
depends on a single dimensionless quantity
\begin{equation}
\chi \equiv \frac{L}{\Lbin} \;, \quad L\equiv bv, \; \Lbin \equiv \sqrt{2G\Mbin a} \;.
\end{equation}
$\chi=1$  corresponds to a distance of closest approach of the field star 
to the binary's center of mass equal to $a$, if the binary were replaced by a single point mass.
We adopt the dependence of $C$ on $\chi$ displayed in Figure~1 of \cite{SesanaHM2006}, 
which we find to be reasonably well approximated (for a circular, equal-mass binary) by
\begin{equation}  \label{eq:chi_approx}
C(\chi) \approx \left\{ \begin{array}{ll}
  1.05 - 1.5\chi^2+21.67\chi^3-25\chi^4 \;, & \chi<0.6 \\
  1.95[1 + 7(\chi-0.6)]\, \mathrm{e}^{-7(\chi-0.6)} \;, & \chi\ge 0.6 \end{array} \right.
\end{equation}

We have ignored the dependence of the scattering cross section on the 
relative orientation of the binary's orbit and that of the incoming star.
The change in binary hardness in one encounter is then given by
\begin{equation}  \label{eq:deltaa_one}
\Delta \left(\frac{1}{a}\right) = \frac{2m_\star}{\Mbin\,a}\, C(\chi) \equiv {\cal H}(\chi)\;.
\end{equation}

Consider for the moment a model in which field stars are drawn from a
homogeneous background with isotropically distributed velocities 
having a single magnitude $v$.
Then the hardening rate is
\begin{subequations}
\begin{eqnarray}  \label{eq:hardening_rate_uniform}
s &\equiv& \frac{d}{dt} \left(\frac{1}{a}\right) =
\frac{\rho v}{m_\star} \int_0^\infty 2\pi b\,db\, {\cal H} (\chi) = 
\frac{G \rho}{v} H_0 \;, \\
H_0 &\equiv& \int_0^\infty 8\pi\,C(\chi)\,\chi\, d\chi \;. \label{eq:H0}
\end{eqnarray}
\end{subequations}

Integrating Equation~(\ref{eq:H0}) using the expression~(\ref{eq:chi_approx}) for 
$C(\chi)$ yields $H_0\approx 18.5$,
almost the same as the value given by \citet{Quinlan1996} in the hard-binary limit.

Next, we derive similar expressions for the hardening rate in the more realistic case
where the distribution function of unbound stars has the form $f(E, L)$.
We take into account that stars interact with the binary once per radial period 
$\trad$, and that the number density of stars in $\{E,L\}$ space is related to 
the phase-space mass density by $dN = 8\pi^2m_\star^{-1} 
\trad(E,L)\, L\, f(E,L) dE\,dL$. 
Then
\begin{eqnarray}
s &=& \int_{\Phi_0}^0 dE \int_0^{L_\mathrm{circ}(E)}\!\!\!\! dL\, 8\pi^2\trad L \frac{f(E,L)}{m_\star} 
\frac{2m_\star\, C(L/\Lbin)}{\Mbin\, a\, \trad} \!\!\!  \nonumber \\
  &=& 4\pi G \int_{\Phi_0}^0 dE 
  \int_0^{L_\mathrm{circ}/\Lbin}\!\!\!\! f(E, \chi\Lbin)\, 8\pi \chi C(\chi) d\chi
  \label{eq:hardening_rate_fEL} .
\end{eqnarray}

Here $\Phi_0$ is the lowest energy of an orbit unbound to the 
central object (equal to the depth of potential well produced by the stars), 
and $L_\mathrm{circ}(E)$ is the angular momentum of 
a circular orbit of energy $E$, which is much larger than $\Lbin$ in the limit of a hard binary. 
If the distribution function is isotropic,
the hardening rate is given simply by \citep[e.g.,][Equation~11]{Merritt2006}
\begin{equation}  \label{eq:hardening_rate_fE}
s_\mathrm{iso} = 4\pi G\, H_0 \int_{\Phi_0}^0 f(E) \,dE \;.
\end{equation}

In the spherical case, this is the rate expected for the ''full-loss-cone'' regime, 
in which the initially isotropic distribution of orbits in angular momentum remains fixed 
(i.e.\ the loss cone is repopulated efficiently enough that we can neglect its depletion).
In the nonspherical case, the rate would be roughly the same if we keep the orbit 
population fixed, even though any individual particle may precess into and out of the loss cone 
due to regular changes of angular momentum in addition to random perturbations. 
Alternatively, we could estimate the full-loss-cone hardening rate
by taking the expression for the hardening rate in a homogeneous isothermal
background, $s=G\rho H/\sigma$, with $H\approx 15$ \citep{Quinlan1996,SesanaHM2006}, 
and substituting the values of $\sigma$ and $\rho$ computed at the binary's
radius of influence (say), which yields a similar number.
For our models during the late stages of evolution ($t\gtrsim 30$),
Equation~(\ref{eq:hardening_rate_fE}) yields $s\approx 18\pm 2$,
almost independent of $N$ and geometry.
This value is consistent with the hardening rates observed in our lowest-$N$ simulations
(Figure~\ref{fig:hardening_rate}),
but is several times {\it higher} than that of the simulations with $N=10^6$, 
for all three geometries: additional evidence that even our triaxial models 
are far from being in the full-loss-cone regime.

Another way to justify the conclusion just reached is by calculating 
$s$ directly from the \Nbody 
discrete distribution function, by summing the contributions to energy exchange 
(\ref{eq:deltaa_one}) for each particle with angular momentum $L_i$ during its 
orbital period ${\trad}_{,i}$:
\begin{equation}  \label{eq:hardening_rate_nbody}
s_{N\mathrm{-body}} = \sum_{i=1}^N \frac{2m_i}{\Mbin a\, {\trad}_{,i}} C(L_i/\Lbin) \;.
\end{equation}

Computed in this way, the predicted hardening rates for all three series of models 
agree quite well with those measured in the simulations (Figure~\ref{fig:hardening_rate}, 
bottom panel). We then artificially randomized the directions of the 
velocities of all the stars, leaving their magnitudes unchanged, 
thus creating an isotropic stellar system, which we integrated forward in time.
(This was only done for a spherical system, since it would break the self-consistency 
of the nonspherical ones). The measured hardening rate was found to 
briefly jump to the full-loss-cone value for 1--2 time units before returning to the 
previous value. 

\begin{figure}
$$\includegraphics{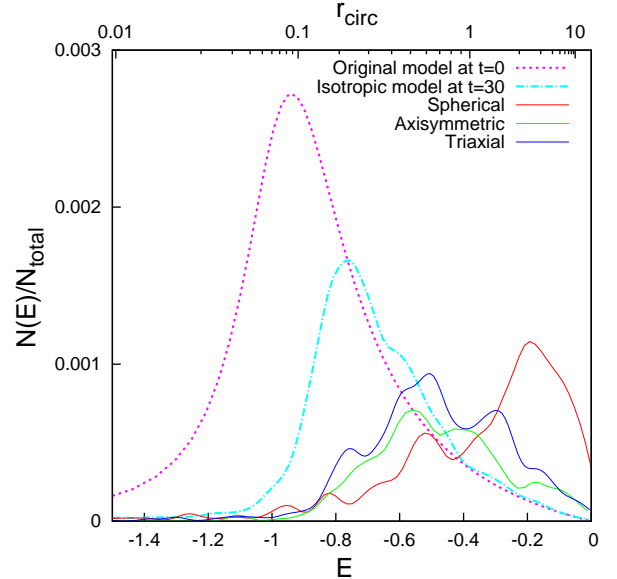}$$
\caption{
Population of stars in the loss cone (with angular momenta $L<\Lbin\equiv \sqrt{2G\Mbin a}$) 
at $t=30$ ($a=1/250$) as a function of energy.
Solid lines: the actual snapshots of spherical (red), axisymmetric (green), and triaxial (blue) 
models with $10^6$ particles; cyan dot-dashed line: population in an isotropic model with 
the same density profile (i.e., full loss cone); dashed magenta line: same in the original 
 model (before the formation of a hard binary). 
} \label{fig:loss_cone_population}
\end{figure}

The degree of loss-cone depletion can be quantified by plotting the actual population 
of particles on loss-cone orbits as a function of energy and comparing it with the isotropic case.
Figure~\ref{fig:loss_cone_population} shows that indeed the population of low angular momentum 
stars is much depleted compared with that of an isotropic model. For the S model, however, 
there is an {\it excess} of stars with low binding energies in the loss cone.
These are the stars that have interacted once with the binary 
but that remained bound by the galactic potential, making them
candidates for the secondary slingshot process \citep{MilosMerritt2003b}. 
Interestingly, those stars are absent in A and T models, presumably because angular momentum 
is not conserved and they precess away from the loss cone.

\subsection{Theoretical estimates}  \label{sec:draining_model}
The analysis presented above showed that there is only a 
moderate difference in the degree of loss-cone depletion between the S, A and T models. 
The goal of this section is to explain that result, and to show that for the present simulations,
we would not expect to be able to reliably distinguish between collisional and collisionless 
loss-cone refilling processes. 

For this purpose, we develop a simplified model of collisionless draining of 
the ``loss region,'' defined as the region of phase space from which orbits of stars in 
a nonspherical potential are able to reach low enough angular momenta that they can 
interact with the binary. 
(By contrast, the ``loss cone'' is the subset of trajectories in the loss region that are
destined to pass near the binary in a single radial period or less.)
This treatment is similar to the orbital draining models considered by \citet{Yu2002}, 
\citet{MerrittPoon2004}, \citet{MerrittWang2005}, and \citet{SesanaHM2007} for a binary 
black hole, or by \citet{VasilievMerritt2013} for a single black hole in an axisymmetric 
galaxy, but is more realistic in that we:
(1) use the actual orbit population of the \Nbody simulations;
(2) take into account the movement of orbits into and out of the loss cone
due to torquing by the mean-field potential;  
and (3) adopt a time-dependent size of the loss region around the central object.

First, we take a snapshot from the actual simulation at $t=30$, when the initial stage of 
rapid evolution of the density profile has finished and a constant hardening rate of
the binary has set in.
We study the orbital population of the model by extracting a sample of $10^5$ particles 
from the snapshot and evolving them in a fixed background potential corresponding to 
the same snapshot (plus one central point of mass $\Mbin$), but represented as 
a combination of smooth functions of radius times spherical harmonic functions of 
the angles $(\theta,\phi)$
(up to $l_\mathrm{max}=6$ in angular harmonics, 
keeping only triaxial or axisymmetric terms as appropriate; see \citet{Vasiliev2013}
for details).
We then follow each orbit for 200 $\trad(E)$, which is long enough to build a meaningful 
distribution of values of the angular momentum at times of pericenter passage 
(see the Appendix for details).
For a given orbit $i$, the probability of having $L^2$ at pericenter below a certain value 
$X$ is found to be well described by a linear function with slope $\mathcal{S}_i^{-1}$:
\begin{equation}  \label{eq:L2prob}
\mathcal{P}(L^2<X) \approx \frac{X-{\Lmin}_{,i}^2}{\mathcal{S}_i} \;.
\end{equation}

A zero value of ${\Lmin}_{,i}$ indicates a truly centrophilic orbit, which can only exist in 
a triaxial potential; however, all orbits with ${\Lmin}_{,i}\lesssim \Lbin$ are ``useful'' 
for loss-cone repopulation. 
The combined mass of these orbits is 6\% (1.5\%) of the entire model for the T (A) case, 
i.e., substantially higher than the mass of the binary.
We assume that values of $L^2$ at subsequent pericenter passages are uncorrelated, which is 
reasonable given that most orbits of interest appear to be chaotic. 

Next, we consider a time-dependent model for binary evolution that includes a depletion 
of orbits in the loss region. 
At each time step, we compute the instantaneous hardening rate according to 
Equation~(\ref{eq:hardening_rate_nbody}), with $C(L_i/\Lbin)$ for each particle being 
averaged over all possible values of angular momentum at pericenter, weighted with 
the probability distribution (\ref{eq:L2prob}):
\begin{subequations}  \label{eq:hardening_rate_drain} 
\begin{eqnarray}
s(t) &\equiv& \frac{d}{dt} \left(\frac{1}{a}\right) =
  \sum_{i=1}^N \frac{2m_i(t)}{\Mbin\, a(t)\, {\trad}_{,i}} 
  \int_0^1 d\mathcal{P}\,C(\chi(\mathcal{P})) = \!\!\!\!\!\!\!\!\!\!\, \nonumber \\
  &=& \sum_{i=1}^N \frac{G\, m_i(t)}{\pi\, \mathcal{S}_i\, {\trad}_{,i}}  H_i(t) \;,
 \\
H_i &\equiv& \int_{\chi_\mathrm{min}}^\infty  8\pi\,C(\chi)\,\chi\, d\chi \;, 
\quad \chi_\mathrm{min} \equiv \frac{{\Lmin}_{,i}}{\Lbin(t)}
  \label{eq:Hi}
\end{eqnarray}
\end{subequations}

The quantity $H_i$ equals the constant $H_0$ (\ref{eq:H0}) for a centrophilic orbit 
($\chi_\mathrm{min}=0$), and tends to zero for orbits with ${\Lmin}_{,i} \gtrsim 2\Lbin$.

Next we need to account for the decrease in the mass $m_i$ associated with each orbit, 
since a star once scattered is assumed not to interact again with the binary. 
To account for this, we relate the evolution of $a$ 
to the mass ejection rate, as described by a dimensionless coefficient 
\begin{equation}  \label{eq:J}
J \equiv \frac{a}{\Mbin} \frac{dM_\mathrm{ej}}{da}
\end{equation}
\citep{Quinlan1996}, and write 
\begin{equation}
\frac{dM_\mathrm{ej}}{dt} = -\sum_{i=1}^N \frac{dm_i}{dt} = J\,\Mbin\,a\,s .
\end{equation}

Identifying each term in this sum with the corresponding term in 
Equation~(\ref{eq:hardening_rate_drain}) allows us to write
the equation for the time evolution of the $m_i$:
\begin{equation}  \label{eq:orbit_mass_decrease}
\frac{dm_i}{dt} = -m_i \frac{G\Mbin\,a(t)\,J H_i(t)}{\pi \mathcal{S}_i\,{\trad}_{,i}}
\end{equation}

\begin{figure}
$$\includegraphics{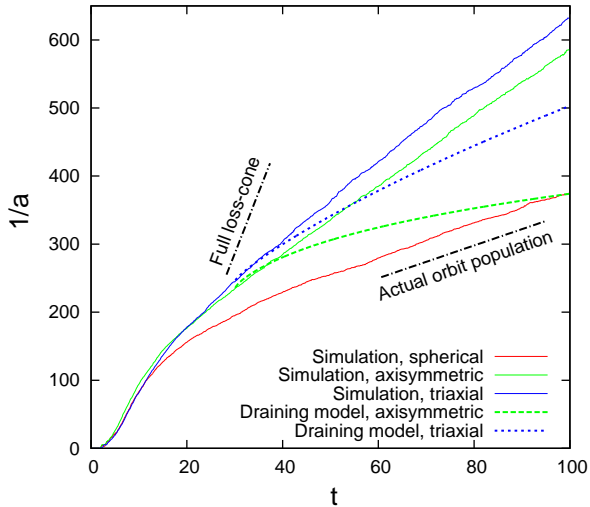}$$
\caption{
Evolution of binary hardness as a function of time. \protect\\
Solid lines: actual simulations with $N=10^6$, from top to bottom: T, A and S models; 
dotted lines: predictions from collisionless draining models, obtained by integrating 
Equations (\ref{eq:hardening_rate_drain}) and (\ref{eq:orbit_mass_decrease}) forward in time, 
starting at $t=30$ with $1/a\simeq 250$. Upper curve is for T model, lower is for A model; 
both are substantially lower than the actual hardening rates due to combined effects 
of draining and relaxation, and comparable to the hardening rate of the S model, which 
is due to relaxation only. 
Dot-dashed lines show the hardening rate expected in the full-loss-cone regime 
(Equation~\ref{eq:hardening_rate_fE}) and the one computed from the actual orbit population 
(Equation~\ref{eq:hardening_rate_nbody}) for the spherical model.
} \label{fig:draining_rate}
\end{figure}

This simplified treatment does not account for the variance in the outcome of 
scattering events with a given impact parameter, 
but is sufficient for the purposes of our estimate.
The value of $J$ is taken from \citet{SesanaHM2006}; it has little dependence on $a$ and 
lies in the range 0.5--1 and we took the former value, which was also found in the \Nbody 
simulations of \citet{MilosMerritt2001}. A higher value of $J$ would decrease the late-time 
hardening rate, as the orbits would be depleted faster.
Equations (\ref{eq:hardening_rate_drain}) and (\ref{eq:orbit_mass_decrease}), 
together with the initial conditions $a(0)=a_\mathrm{init}$, $m_i(0)=1/N$ and the 
coefficients ${\Lmin}_{,i}, \mathcal{S}_i, {\trad}_{,i}$ derived from the orbit analysis, 
describe the time-dependent evolution of binary hardness in the presence of orbital draining 
due to non-conservation of angular momentum in a nonspherical background potential.

Figure~\ref{fig:draining_rate} shows the predicted evolution of $1/a$ in
models T and A starting at $t=30$ and $1/a=250$, compared with 
the actual evolution of $1/a$ in the simulations with $N=10^6$.
We also show, for the spherical model, the evolution rate expected for the full-loss-cone 
case and for the actual population of loss-cone orbits.
Two results are apparent:
(1) binary hardening rates predicted by our (collisionless) machinery in the
A and T models fall below the actual rates, but only by modest factors.
In other words, collisional replenishment of the orbits is still contributing
to the evolution of the binary in these models.
(2) The hardening rates in the A and T models -- both predicted and actual -- fall
substantially below the full-loss-cone rate.
In other words, replenishment of orbits by torquing due to the nonspherical
potential is not efficient enough to keep the loss-cone orbits fully occupied.

Figure~\ref{fig:draining_rate} suggests that -- even for this large number of particles --
we cannot reliably discern the relative contribution of collisional and collisionless 
processes to the hardening rate.
A consequence of this result is that it is difficult to extrapolate our results 
to the much higher $N$ values relevant to real galaxies. 

At the same time, our simplified model suggests that orbital draining can sustain 
a hardening rate that is several times below that of a full loss cone for a fairly 
long time in the triaxial case, as the total mass of stars on centrophilic orbits is 
a few times larger than $\Mbin$.
The situation is less clear for the axisymmetric models: on the one hand, there are no 
``genuinely centrophilic'' orbits in this case; on the other hand, the reservoir of 
orbits that have low enough $\Lmin$ to be able to interact with the binary is still 
much larger than the volume of the loss cone in the spherical case. The simplified 
calculation above predicts that the draining rate of this loss region is merely a factor 
of two lower than in the triaxial case, but that it may depend strongly on time.

\subsection{Discussion}  \label{sec:discussion}
The rate of repopulation of loss-cone orbits is determined by a combination of collisional 
effects (due to gravitational encounters) and collisionless effects (due to torquing by a 
nonspherical potential). 
The former are well studied in the context of standard loss-cone theory.
Approximate Fokker--Planck models for the binary evolution in the spherical geometry 
\citep{MilosMerritt2003b,MerrittMS2007} predict that the hardening rate should scale 
as $s \propto N^{-1}$ for large $N$ (when the system is fully in the empty-loss-cone 
regime), and the numerical simulations cited above show substantial $N$-dependence 
(although less steep than the prediction) for $N \gtrsim 10^5$. 

The second mechanism of loss-cone repopulation, which can torque orbits from a much 
larger ``loss region'' into the loss cone, is  collisionless and therefore does not 
depend on $N$. Still, in our \Nbody integrations, we have observed a substantial 
$N$-dependence in both the axisymmetric and triaxial cases. 
A simple model for draining of the loss region 
found that this mechanism could account for about one-half of the 
hardening rate seen in the simulations; the remainder would be attributed to 
$N$-dependent collisional effects. 
To reliably drive the latter  below the expected rate of collisionless loss-cone 
repopulation would require a still much larger value of $N$ -- difficult to achieve
with existing algorithms and hardware. 

An additional complication arises from the complex interplay between 
collisional and collisionless factors: the size of the loss region, from which the orbits 
can be torqued into the loss cone due to collisionless effects, is much larger in 
nonspherical than in spherical systems, 
and hence can be more readily repopulated by {\it collisional} relaxation. 
For instance, \citet{VasilievMerritt2013} have shown that in a steady state, the rate 
of loss-cone repopulation for a single SMBH in an axisymmetric galaxy is a few times 
higher than in the corresponding spherical system.
This makes it even more difficult to design an \Nbody simulation in which the collisional 
effects would be negligible, as they are expected to be in many real galaxies.
While we certainly expect there to exist an effective lower limit, as a function of
$N$, on the binary hardening rates in nonspherical galaxies, we are unable to make
a precise statement concerning how low that rate might be.
Our results suggest only that we cannot approach that rate with currently available
algorithms and computing hardware.

The simple draining model considered in this paper does not account for 
a number of other processes that may be effective in the \Nbody simulations
or in real galaxies: Brownian motion of the binary \citep{Merritt2001}, 
the ``secondary slingshot'' \citep{MilosMerritt2003b}, 
time-dependent perturbations to stellar orbits even far from the binary \citep{KandrupSTB2003}, 
changes in the stellar density profile due to continuous ejection of stars \citep{MilosMerritt2001}, 
possible long-term changes in degree of triaxiality \citep{MerrittQuinlan1998}, among others.
Clearly, a still more elaborate model, combining both collisionless and collisional processes, 
is desired to better understand the dynamics of binary SMBHs in galactic nuclei. 

Even in the context of models like ours, binary hardening rates will vary as a function
of the degree of velocity anisotropy (i.e., the detailed orbital population),
the radial density profile, the shape of the nuclear isodensity contours, etc.
While exploring the full range of such variation is beyond the scope of this paper,
we can make some general remarks.
Observed galaxies exhibit a variety of nuclear density profiles, from nearly flat
cores to nuclear star clusters having $\rho\sim r^{-2}$.
At larger radii the density is almost always well fit by a Sersic or Einasto function.
Varying the nuclear density profile would certainly affect the early hardening rate
of a binary.
As discussed above (Section 2.3), during the formation of a ``hard'' binary,
the initial density cusp is converted into a shallower profile
as stars bound to the binary are ejected.
The same will be qualiatively true for any initial nuclear density profile, although
one expects some dependence of the final profile on the initial profile
\citep{MilosMerritt2001,Khan2012}.
A velocity distribution that is more or less anisotropic would also lead to higher
or lower binary hardening rates, at least initially.
With regard to changes in the shape of the model,
the results obtained here suggest that in nonspherical geometries,
even fairly radical shape changes (axisymmetric $\rightarrow$ triaxial)
have only modest consequences for binary hardening rates, and we expect the
same to be true in nonspherical models with axis ratios different from those considered
here.

Our results present an interesting contrast to those of other recent studies
based on similar techniques.
\citet{KhanHBJ2013} integrated spherical and axisymmetric models created initially in equilibrium,
with density profiles and black hole masses essentially the same as in our models.
While their binary hardening rate for a spherical model with $N=10^6$ is comparable to ours, 
their flattened models have a much higher hardening rate than ours, 
even exceeding our estimate for the full-loss-cone case (\ref{eq:hardening_rate_fE}).
We also observed much less difference between the spherical and nonspherical 
models up to $N=10^6$. The reasons for these differences are unknown to us;
however, at face value, our results call into question the robustness of the
conclusion reached by those authors that the final-parsec problem is ``solved'' in axisymmetric galaxies.

On the other hand, triaxial models formed by a bar instability in a rotating galaxy 
\citep{BerczikMSB2006}, or by mergers of two galaxies \citep{KhanJM2011, PretoBBS2011}, 
have shown essentially no dependence of the hardening rate on $N$.
We speculate that this difference with our results might be due to the rotation of
those models, to non-stationary clumpy structures in the case of the merger remnants, 
or to some other factor. 
More detailed study of the orbit populations could shed light on this mystery. 
 
\citet{MerrittPoon2004} (MP04) analyzed the orbital populations in self-consistent,
triaxial models of nuclei containing central SMBHs.
In the context of scale-free models with a steep, $\rho\sim r^{-2}$ density profile, they
argued that collisionless feeding rates might be high enough to ensure 
coalescence of massive binaries in less than $10$ Gyr, even in galaxies where the
fraction of centrophilic orbits was small.
The models of MP04 were both extreme (a steep density profile,
maximal triaxiality) and idealized (scale-free, fixed potential).
We attempted to verify their conclusions by creating a non-scale-free model with similar 
central properties (a steep density cusp and strong triaxiality) and an SMBH mass
of $10^{-2} M_\mathrm{gal}$; in this model the region extending to a few influence radii 
was still well inside the break radius. 
This model had $\sim 15\%$ of its mass on chaotic/centrophilic orbits, 
somewhat more than in the models presented above. 
The time-dependent draining rate was then computed as described above,
 assuming an initial binary separation of $0.5 a_\mathrm{hard}$. 
Our results for $a(t)$ were not precisely the same as those in MP04; 
in particular, the hardening rate was found to drop more rapidly with time, 
$(a_\mathrm{hard}/a) \sim t^{0.65}$.  
But the value of $(a/a_\mathrm{hard})^{-1}$ after a time corresponding to several Gyr
was $\sim 10^3$, i.e., more than enough to ensure GW coalescence 
on Gyr timescales.

This comparison highlights an important point: even a modest rate of binary
evolution (due to stellar-dynamical interactions) can result in a separation
small enough for GW emission to induce coalescence in less than a Hubble time.
The time for a circular binary's orbit to evolve, from $a=a_0$ to $a=0$, 
due to GW emission is 
\begin{eqnarray}\label{Equation:DefinetGW}
t_\mathrm{GW}  &\equiv& t(a=0) - t(a=a_0)=\frac{5}{256}\frac{c^5a_0^4}{G^3M_1M_2M_\mathrm{bin}}\\
\label{Equation:tGWbinary}
&\approx& 5.7\times 10^{6}
\frac{(1+q)^2}{q}
\left(\frac{a_0}{10^{-2}\,\mathrm{pc}}\right)^4
\left(\frac{M_\mathrm{bin}}{10^8\,M_\odot}\right)^{-3}\,\mathrm{yr}
\nonumber
\end{eqnarray}
with $q\equiv M_2/M_1\le 1$ \citep[][Equation~(4.241)]{MerrittBook}.
This time is less than $10$ Gyr if
\begin{equation}
\left(\frac{a_0}{a_\mathrm{hard}}\right)^4 \lesssim \left(1.3\times 10^{14} \mathrm{yr}\right)
\;\frac{(1+q)^6}{q^3}\frac{G^3M_\mathrm{bin}^3}{c^5 r_m^4} 
\end{equation}
i.e. if
\begin{equation}
\frac{a_0}{a_\mathrm{hard}} \lesssim 0.03 \frac{(1+q)^{3/2}}{q^{3/4}}
\left(\frac{M_\mathrm{bin}}{10^8 M_\odot}\right)^{3/4}
\left(\frac{r_\mathrm{m}}{10\; \mathrm{pc}}\right)^{-1}
\end{equation}
or in the units of our $N$-body models ($a_\mathrm{hard} \approx 0.01$,
$q=1$)
\begin{equation}
a_0 \lesssim 10^{-3} \left(\frac{M_\mathrm{bin}}{10^8 M_\odot}\right)^{3/4}
\left(\frac{r_\mathrm{m}}{10\; \mathrm{pc}}\right)^{-1} .
\end{equation}
Assuming that we have nearly reached the asymptotic (large-$N$) limit in
our simulations, Figure 1 suggests that the binaries in many galaxies would
indeed be able to reach coalescence in a Hubble time.

\section{Conclusions}

We carried out direct \Nbody integrations of binary supermassive black holes
in spherical, axisymmetric, and triaxial models of galaxies, constructed initially as 
equilibrium models. 
Our integrations with particle numbers up to $N=10^6$ demonstrated that in all three 
geometries considered, the binary hardening rate $s\equiv d(1/a)/dt$ ($a=$ binary 
semimajor axis) at late times does depend on $N$ and is several times below the rate 
computed assuming a full loss cone. 
The difference in hardening rates between the three models was quite modest -- 
within a factor of two even for the simulations with largest $N$ -- and only in 
the largest-$N$ case was there a noticeable difference in $s$ between the 
axisymmetric and triaxial geometries.

To assist in understanding these results, we computed the expected hardening 
rates based on known results from three-body scattering experiments, together
with the distribution of particles in energy and angular momentum in the \Nbody models.
These predictions were found to agree well with the hardening rates 
obtained in the actual simulations. 
We also estimated, using a simple model for collisionless draining of orbits
in the ``loss region'' (the collection of orbits that are able to reach the 
binary's interaction sphere),
the contribution of nonspherical torques to the rate of loss-cone repopulation, 
and we found it to be below or comparable to the contribution from collisional effects, 
even for the highest-resolution simulation of our set. 

Based on these results, we argued that 
in order to reach a regime that is characteristic of massive galaxies
(in which collisional effects are believed to be negligible),
substantially higher values of $N$ might be needed in the simulations.
Until this is done, it is premature to state that the final-parsec problem in
gas-free galaxies is ``solved'' by assuming nonspherical geometries.

\bigskip
This work was supported by the National Science Foundation under grant no. AST 1211602 
and by the National Aeronautics and Space Administration under grant No. NNX13AG92G.
Part of this work was done at the Alajar meeting\footnote{http://members.aei.mpg.de/amaro-seoane/ALM13} 
and in the Henri Poincar\'e institute in Paris.
Computations were performed on the GPUs of the CITA Sunnyvale cluster, as well as the ARC 
supercomputer at the SciNET HPC Consortium. SciNet is funded by the Canada Foundation 
for Innovation under the auspices of Compute Canada, the Government of Ontario, 
Ontario Research Fund Research Excellence, and the University of Toronto.
We thank John Dubinski for assistance with the cluster.


\appendix
The analysis of draining rates in Section \ref{sec:draining_model} necessitated an estimate
of the minimum angular momentum $\Lmin$ attained by a given orbit in the smooth potential. 
Of course, no orbit can reach zero angular momentum on any finite time interval 
(unless it is specially arranged to do so), but one can nevertheless estimate whether 
there is a positive lower limit on $\Lmin$, or whether it is compatible with being zero, 
by the following procedure.

We record the values of the squared angular momentum at pericenter passages 
$L_{\mathrm{peri},k}^2$ ($k=1..N_\mathrm{peri}$) and sort them in ascending order. 
As discussed in the text, it happens that the distribution 
usually follows a linear trend at low $L^2$.
We therefore fit a linear regression with and without 
a constant term: 
\begin{equation}
L_{\mathrm{peri},k}^2 = \Lmin^2 + s\, \frac{k}{N_\mathrm{peri}} + \delta_k =
s'\, \frac{k}{N_\mathrm{peri}} + \delta_k'\;, \quad  
k=1..N_\mathrm{fit}, \;\; N_\mathrm{fit}=0.1\,N_\mathrm{peri} .
\end{equation}
Here, $s,\Lmin^2$ and $s'$ are the coefficients of two- and one-parameter fits, 
and $\delta_k$ and $\delta_k'$ are the corresponding residuals. 
We assign the intrinsic dispersion of the values $L_{\mathrm{peri},k}^2$ from the condition 
that $\chi^2$ per degree of freedom is unity in the two-parameter fit: 
$\sigma^2 \equiv \left(\sum_{k=1}^{N_\mathrm{fit}} \delta_k^2 \right)/(N_\mathrm{fit}-2)$. 
Next we compare the statistical significance of the fits to find out whether to prefer the 
one-parameter fit (describing a centrophilic orbit) over the more general two-parameter one. 
The difference $\Delta\chi^2$ between the one-parameter and two-parameter fits is given by
\begin{equation}
\Delta\chi^2 = (N_\mathrm{fit}-2) \left(
  \frac{\sum_{k=1}^{N_\mathrm{fit}} {\delta'}_k^2}{\sum_{k=1}^{N_\mathrm{fit}} \delta_k^2} - 1 
\right) \;.
\end{equation}
Of course, the residuals in the one-parameter fit are always greater than in the two-parameter 
fit, but if they are ``not too much'' greater then we accept the hypothesis that $\Lmin^2=0$. 
More quantitatively, we accept the one-parameter fit if it is less than $3\sigma$ away from 
the two-parameter fit, i.e., if $\Delta\chi^2<\Delta\chi^2_{3\sigma}\equiv 11.8$, the latter 
value being the $3\sigma$ deviation for a $\chi^2$ distribution with two degrees of freedom.
Usually if this hypothesis is rejected (i.e., the orbit is labeled centrophobic), it is 
at the level of significance of many hundreds or thousands of $\sigma$.
Finally, we take the values of $\Lmin^2$ and $s$ (or $s'$) from the adopted regression 
to estimate the probability of having a given value of $L^2_\mathrm{peri}$ 
(Equation~\ref{eq:L2prob}).

\end{document}